\documentclass[twoside,preprint,11pt]{article}

\usepackage[hyperref]{jmlr2em}

\usepackage{amsmath,amssymb,amsfonts}
\usepackage{algorithmic}
\usepackage{graphicx}
\usepackage{textcomp}

\usepackage{booktabs}
\usepackage{xurl}

\usepackage{orcidlink}

\usepackage{marvosym}

\usepackage{lastpage}
\jmlrheading{}{2025}{}{}{}{}{Mays Al-Azzawi, Dung Doan, Tuomo Sipola, Jari Hautamäki and Tero Kokkonen}

\ShortHeadings{Red Teaming with Artificial Intelligence-Driven Cyberattacks}{Al-Azzawi, Doan, Sipola, Hautamäki and Kokkonen}
\firstpageno{1}

\begin{document}

\title{Red Teaming with Artificial Intelligence-Driven Cyberattacks: A Scoping Review}
\footnotetext[1]{An earlier version of this paper was presented at the 2024 World Conference on Information Systems and Technologies (WorldCIST'24) and was published in its proceedings \emph{Good Practices and New Perspectives in Information Systems and Technologies} (pp.~129--138) by Springer Nature~\citep{alazzawi2024}.}

\author{\name Mays Al-Azzawi \orcidlink{0009-0008-5693-1388} \email mays.al-azzawi@student.jamk.fi\AND
Dung Doan \orcidlink{0009-0004-4646-7404} \email dung.doan2@student.jamk.fi\AND
Tuomo Sipola \orcidlink{0000-0002-2354-0400} (\Letter) \email tuomo.sipola@jamk.fi\AND
Jari Hautamäki \orcidlink{0000-0002-0519-5594} \email jari.hautamaki@jamk.fi\AND
Tero Kokkonen \orcidlink{0000-0001-9988-6259} \email tero.kokkonen@jamk.fi\\
\addr Institute of Information Technology\\
Jamk University of Applied Sciences\\
PO Box 207, FI-40101, Jyväskylä, Finland
}
\footnotetext[2]{\Letter\ Corresponding author: Tuomo Sipola (e-mail: tuomo.sipola@jamk.fi)}

\editor{Preprint submitted to arXiv.}

\maketitle

\begin{abstract}
The progress of artificial intelligence (AI) has made sophisticated methods available for cyberattacks and red team activities. These AI attacks can automate the process of penetrating a target or collecting sensitive data. The new methods can also accelerate the execution of the attacks. 
This review article examines the use of AI technologies in cybersecurity attacks. It also tries to describe typical targets for such attacks. 
We employed a scoping review methodology to analyze articles and identify AI methods, targets, and models that red teams can utilize to simulate cybercrime. 
From the 470 records screened, 11 were included in the review. Various cyberattack methods were identified, targeting sensitive data, systems, social media profiles, passwords, and URLs. 
The application of AI in cybercrime to develop versatile attack models presents an increasing threat. Furthermore, AI-based techniques in red team use can provide new ways to address these issues.
\end{abstract}


\section{Introduction}
\label{sec:introduction}
 
The possibility of artificial intelligence (AI) simulating human behavior has emerged as a significant cybersecurity threat. As AI technologies continue to advance, so do their potential applications in executing cyberattacks. These technologies can generate highly convincing phishing emails, malicious URLs, and other deceptive content, making it increasingly difficult for individuals and security systems to distinguish between legitimate and malicious activities. Although there have only been a few reported instances of AI-driven cyberattacks, the feasibility and frequency of such attacks are expected to rise as AI capabilities and attack methodologies evolve.

Training in a safe environment with realistic threat technologies is an essential step in achieving credible cyber defense.
The concept of red teaming originates from the military domain, where it is used as a strategy to role-play adversaries and assess vulnerabilities within systems~\citep{longbine2008}. The term red team is also derived from widely used military symbols, such as APP-6 by NATO and MIL-STD-2525 by the U.S. Department of Defense, where the hostile (and suspect) identity is indicated with a red color~\citep{APP-6,MIL-STD-2525}. In the context of cybersecurity, the U.S. National Institute of Standards and Technology (NIST) defines a red team as follows:~\textit{``A group of people authorized and organized to emulate a potential adversary’s attack.''}~\citep{NIST} 
The red teams improve enterprise security by demonstrating the impacts of successful attacks~\citep{NIST}. 
In the context of cybersecurity, the term red team is used in cybersecurity exercises and in security testing. In cybersecurity exercises, red teams (RT) simulate the threat actors of the exercise scenario by executing cyber-attacks against blue teams (BT), which are defending their assets~\citep{Kokkonen_2018,Brynielsson_2016,Wilhelmson_2014,Mitre_2014,Sommestad_2012}. In security testing, the red team is the group of security testers. 


The advancement of AI capabilities, including machine learning, deep learning, and large language models, has opened new avenues for enhancing red teaming activities. AI can significantly improve the efficiency and effectiveness of red teaming by automating various aspects of attack planning and execution. For instance, AI can be used to discover zero-day vulnerabilities~\citep{ali2022}, craft phishing campaigns~\citep{hazell2023}, and generate highly realistic social engineering attacks~\citep{yu2024}. The use of language models in particular has shown great potential in creating convincing phishing emails and messages in multiple languages, increasing the reach and impact of these attacks.

AI red teaming can be understood as an activity from two different perspectives. Several large technology companies utilize red teaming to expose weaknesses and vulnerabilities in their systems~\citep{smith2020case,Zhou2020}. 
Another aspect is the use of AI to carry out attacks, which can be targeted against technical systems. On the other hand, in social engineering-type attacks, AI is used as a stepping stone to advanced persistent threat (APT) attacks by searching for suitable victims that can be targeted by AI-generated ghost messages~\citep{ghafir2015advanced,renaud2023chatgpt}. The advantage of AI specifically in such attacks is the ability to enable mass attacks using phishing techniques to open attack vectors to multiple targets instead of manual attacks. For example, AI-generated phishing messages in target language create persuasive attack vectors.
AI-based solutions are built to make operations more effective. Automating the process of planning attacks for automated cybersecurity testing scenarios could save time and effort~\citep{yuen2015}. 
As new artificial intelligence technologies have become more prevalent, automation is easier to implement, although its impact on work and society should be studied~\citep{wang2019}.

In order to investigate the use of AI for cyberattacks for red teaming, we carried out a scoping review. To examine how AI can be used for cyberattacks, red team actions, and hacking, our research questions were the following: 

\begin{itemize}
\item \emph{RQ1:} What AI attack methods are there? 
\item \emph{RQ2:} What are the targets of such attacks? 
\end{itemize}

The results were first presented in a conference paper~\citep{alazzawi2024}, which has now been extended with revised introduction, additional related literature and clarified results. 
This article is structured so that related literature is described in Section~\ref{sec:related}. The scoping review methodology used in this reasearch is outlined in Section~\ref{sec:methodology}, which also includes a flow chart of the procedure. Next, Section~\ref{sec:results} presents the results, including two tables that summarize the found literature. Section~\ref{sec:discussion} discusses the offensive and defensive tactics revealed by the scoping review. Finally, a conclusion is provided in Section~\ref{sec:conclusion}. 

\section{Related literature}
\label{sec:related}

The use of AI has been the focus of recent research literature. 
Several AI-driven techniques can be used by the red teams to bypass cyber protections. 

Cracking passwords is a familiar yet effective activity to penetrate protections. 
\cite{hassan2023study} describe password guessing and cracking techniques based on AI, and identify three attack types for this purpose: brute-force, guessing, and stealing. 
Furthermore, PassGAN exemplifies an AI-driven attack capable of generating numerous effective password guesses. The impact of AI-powered hacks can cause damage, diminish trust in companies, and cause systemic failures by compromising data confidentiality and integrity~\citep{hassan2023study}.

Offensive cyber actors can use machine learning algorithms to enhance phishing attempts, making the messages harder for cybersecurity systems to detect. 
Red teams have traditionally constructed phishing URLs using random segments. 
The AI-based DeepPhish generates new phishing URLs by learning patterns from past successful phishing URLs~\citep{bahnsen2018deepphish}.
Furthermore, \cite{jackson2023} presents a systematic review of AI-based phishing techniques. 

AI frameworks play a crucial role in predicting cyberattacks through various applications such as breach risk assessment, spam filtering, fraud detection, and behavioral analysis. These frameworks leverage machine learning techniques to analyze data patterns and enhance organizational insights, utilizing approaches like pattern recognition, data mining, and neural networks to detect anomalies in organizational activities and processes effectively~\citep{meduri2024evaluating}.

Artificial intelligence holds great promise for enhancing security measures, yet concerns persist over its reliability and fairness. AI algorithms, often the vague in their decision-making processes, pose challenges in understanding and addressing errors or biases they may introduce, especially when trained on biased data, potentially leading to discriminatory outcomes, to mitigate these risks, transparency and accountability in AI systems are essential. Techniques like explainable AI (XAI) aid in making AI decisions understandable, fostering trust, and enabling effective error identification and correction. Moreover, ensuring diverse and unbiased training data, coupled with rigorous performance testing, is crucial to minimizing errors and biases in security applications. Regulatory oversight plays a critical role in ensuring AI deployment aligns with ethical norms, addressing concerns such as privacy breaches and bias perpetuation, thus safeguarding against unintended consequences and promoting responsible AI innovation~\citep{bharadiya2023ai}.

The defensive nature of AI has also been acknowledged in literature. Employing machine learning and other advanced techniques, AI can analyze vast amounts of data to identify and prevent cyberattacks in real time. This is a significant improvement over traditional methods, which often rely on manual work, predefined rules, and struggle with novel threats. AI's ability to learn and adapt makes it adept at recognizing a new generation of threats, keeping defenders one step ahead of constantly evolving cybercriminals~\citep{sarker2021ai}.
This is echoed by \cite{prasad2023ai}, who argue that organizations, businesses, and governments need cyber security solutions using machine learning to allow analysts to focus on interpreting the results and creating techniques against cybercrime. 

\section{Methodology}
\label{sec:methodology}

We used the scoping review method~\citep{munn2018systematic} to query the Finna\footnote{https://janet.finna.fi/} library database and Google Scholar\footnote{https://scholar.google.com/} for academic articles. The following keywords were used in the queries: {‘defensive mission’, ‘AI-enabled cyber operations’, ‘AI-augmented cyber defenses’, ‘national defense postures’, ‘poisoning attacks’, ‘offensive cyber operations’, ‘Cyber activities’, ‘AI cyber operations’, ‘AI cyber defense’, ‘AI cyber attack’, ‘AI red teaming’, ‘AI-enabled cyber campaigns’, and ‘cyber attacks’. 
The queries for data gathering were performed in mid-2023. 
The scoping was divided into identification, screening and inclusion stages. 
The identification stage, after making the database queries, identified 471 articles (and some book chapters). 
During the screening stage, titles and abstracts were screened for suitability, and the timeframe was set to include the years 2015--2023. 
We included articles written in English with available abstracts. 
A more involved analysis of these articles was conducted. This analysis included reading the articles closely and concentrating on the topic at hand to precisely determine their content and classify them as directly relevant to addressing the research questions (RQ1, RQ2). To be included in the final, inclusion stage, the following inclusion criteria were used:

\begin{enumerate}
\item Is there a description of an attack method? (RQ1)
\item How was the attack conducted? (RQ1)
\item What was the cyber-attack methodology used? (RQ1)
\item What were the targets of the attacks? (RQ2)
\end{enumerate}

The inclusion stage yielded 9 articles and 2 books related to the topics relevant to this research, which is 11 studies included in total. 
Subsequently, in the analysis stage of the research, we composed summaries, which also involved addressing the criteria questions more deeply, when applicable. The detailed analysis of the included studies revealed the current state of AI use in red team activities. 
A PRISMA flow chart~\citep{McGowan_2020} detailing the review process is provided in Figure~\ref{fig:scopingreview}. It shows the identification, screening and inclusion stages of the research, and the number of included articles.

\begin{figure}[htbp]
\centering
\includegraphics[width=0.76\textwidth]{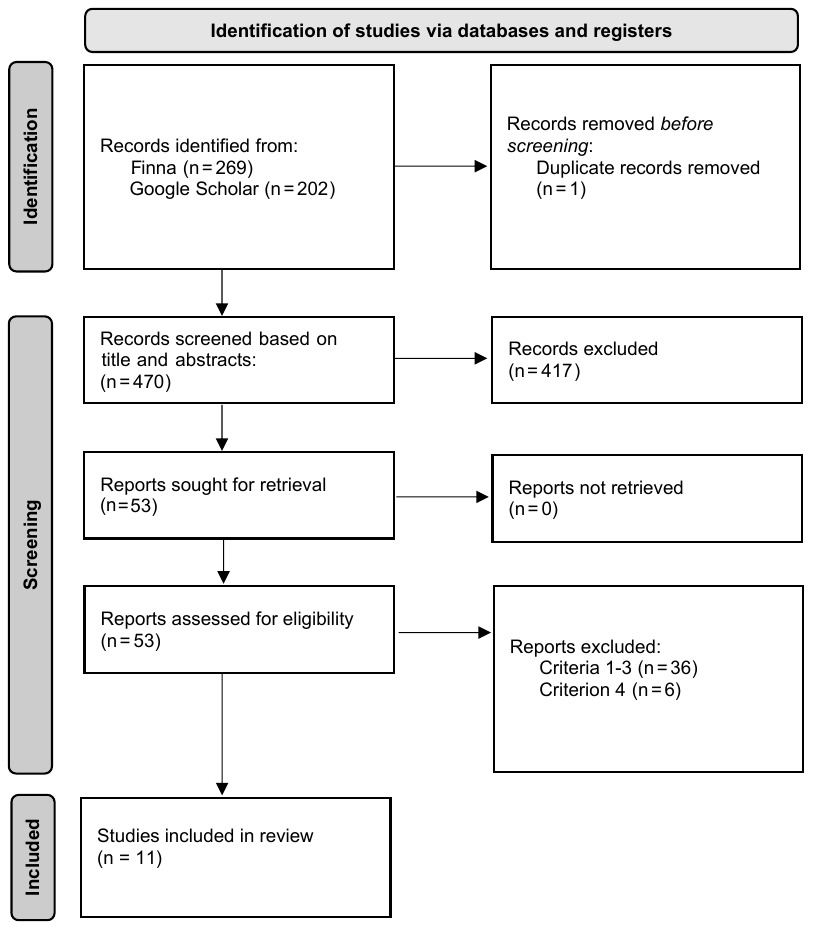}
\caption{Review process.}
\label{fig:scopingreview}
\end{figure}

\section{Results}
\label{sec:results}

\subsection{AI methods used in attacks}

We identified various AI methods used in attacks and also employed in red teaming exercises. As stated, the literature review included articles from 2015 to 2023 to limit the results to recent years. The documented techniques were categorized into four main groups: \emph{Classification methods} , \emph{Regression methods} , \emph{Clustering strategies} , and \emph{Other specific methods}. Each category encompasses specific methods that were observed in the reviewed studies as following: 

\emph{Classification methods:} 
decision tree, 
convolutional neural network (CNN), 
recurrent neural network (RNN), 
long short-term memory (LSTM), 
support vector machine (SVM), 
support vector classification (SVC), 
deep neural network (DNN), 
least squares support vector machine (LS-SVM), 
natural language processing (NLP), 
one-versus-all (OVA), 
double deep Q-network (DDQN), 
advantage actor-critic (A3C) 
regularized least-squares classification (RLSC), 
domain generation algorithms (DGA).

\emph{Regression methods:} 
generative adversarial network (GAN), 
random forest (RF), 
multilayer perceptron (MLP), 
gradient boosting regression trees (GBRT), 
artificial neural network (ANN), 
logistic regression, 
generalized likelihood ratio test (GLRT),

\emph{Clustering strategies:} 
k-means clustering, 
restricted Boltzmann machine (RBM), 
particle swarm optimization (PSO), 
genetic algorithm (GA), 
deep autoencoder (DAE), 
Lagrangian firefly algorithm (LFA),

\emph{Other specific methods:}
nonsymmetric deep autoencoder (NDAE), 
cycle-GAN, combining TensorFlow object detection and a speech segmentation method with convolutional neural network (TOD+CNN), 
k-nearest neighbors (KNN), 
reinforcement learning (RL), 
gray wolf optimization (GWO), 
random weight network (RWN), 
ML-based approach MLAPT, 
software-defined networking (SDN), 
and singular value decomposition (SVD). 

In our comprehensive review of the various methods employed in the analyzed studies, LSTM emerged as the most prominently utilized technique, featuring in five (5) of the reviewed articles. LSTMs are popular because they are good at analyzing sequences of data, making them useful for detecting patterns over time. GANs and SVM were also widely used, each appearing in four studies. GANs can create realistic fake data for training purposes, while SVMs are strong at classifying data into different categories.

Moreover, three of the reviewed articles included CNN, RNN, KNN, MLP, and DNN. The prevalence of CNNs could be due to their pattern recognition capabilities with images and spatial data. RNNs, like LSTMs, handle sequences of data but are simpler. KNN is a straightforward method that classifies data based on the closest examples. MLPs are the basic type of neural network, and DNNs are more complex, able to learn from large amounts of data.

Other methods were referenced only once or twice, suggesting they are either specialized or new to the field.
Table~\ref{tab:AImethods} presents a detailed list of all the AI methods found in the reviewed articles.

\begin{table}
\caption{Methodologies found in the reviewed articles.}
\label{tab:AImethods}
\centering
\scriptsize
\begin{tabular}{l p{1.25cm} p{1cm} p{1cm} p{0.8cm} p{0.8cm} p{0.8cm} p{0.8cm} p{0.8cm} p{0.8cm} c}
\toprule
Ref. & \cite{pistono2016unethical} & \cite{brundage2018malicious} & \cite{kaloudi2020ai} & \cite{king2020artificial} & \cite{truong2020artificial} & \cite{zouave2020artificially} & \cite{YAMIN2021102722} & \cite{wang2022poisoning} & \cite{guembe2022emerging} & $\sum$ \\
\midrule
\multicolumn{11}{c}{Classification}\\
\midrule
Dec. tree & & &   & &   & x &   &   &   & 1 \\
CNN       & & &   & & x & x &   &   & x & 3 \\
RNN       & & & x & &   & x &   &   & x & 3 \\
LSTM      & & & x & & x & x &   & x & x & 5 \\
SVM       & & &   & & x & x &   & x & x & 4 \\
SVC       & & &   & &   & x &   &   & x & 2 \\
DNN       & & & x & &   &   &   & x & x & 3 \\
LS-SVM    & & &   & & x &   &   &   &   & 1 \\
NLP       & & &   & &   & x &   &   &   & 1 \\
OVA       & & &   & &   & x &   &   &   & 1 \\
DDQN      & & &   & &   &   &   & x &   & 1 \\
A3C       & & &   & &   &   &   & x &   & 1 \\
RLSC      & & &   & &   & x &   &   &   & 1 \\
DGA       & & &   & &   & x &   &   &   & 1 \\
\midrule
\multicolumn{11}{c}{Regression}\\
\midrule
GANs      & & &   & &   & x & x & x & x & 4 \\
RF        & & &   & &   & x &   &   & x & 2 \\
MLP       & & &   & & x & x &   &   & x & 3 \\
GBRT      & & &   & &   & x &   &   & x & 2 \\
ANN       & & &   & & x &   &   &   &   & 1 \\
Log. reg. & & &   & &   & x &   &   &   & 1 \\
GLRT      & & &   & & x &   &   &   &   & 1 \\
\midrule
\multicolumn{11}{c}{Clustering}\\
\midrule
k-means   & & & x & &   &   &   &   &   & 1 \\
RBM       & & &   & & x &   &   &   &   & 1 \\
PSO       & & &   & & x &   &   &   &   & 1 \\
GA        & & &   & & x &   &   &   &   & 1 \\
DAE       & & &   & & x &   &   &   &   & 1 \\
LFA       & & &   & & x &   &   &   &   & 1 \\
\midrule
\multicolumn{11}{c}{Other}\\
\midrule
NDAE      & & &   & & x &   &   &   &   & 1 \\
CYCLE-GAN & & &   & &   &   &   &   & x & 1 \\
TOD+CNN   & & &   & &   &   &   &   & x & 1 \\
KNN       & & &   & & x & x &   &   & x & 3 \\
RL        & & & x & &   &   &   &   &   & 1 \\
GWO       & & &   & & x &   &   &   &   & 1 \\
RWN       & & &   & & x &   &   &   &   & 1 \\
MLAPT     & & &   & & x &   &   &   &   & 1 \\
SDN       & & &   & &   &   &   & x &   & 1 \\
SVD       & & &   & &   &   &   & x &   & 1 \\
\bottomrule
\end{tabular}
\end{table}

\subsection{Attack targets}

There are several common categories targeted by attackers. These targets vary widely but can generally be categorized into a few key areas, as detailed in Table~\ref{tab:targets}. Those targets identified in this study include the following five categories:

\begin{itemize}
\item General data: This includes personal data, health data, financial data and government data. Four reviewed articles included this category as a target, making it the most frequent~\citep{brundage2018malicious,truong2020artificial,zouave2020artificially,YAMIN2021102722}. The high frequency of attacks on these data types underscores their value and the significant impact their compromise can have on individuals and institutions.
\item URLs: URLs were another common target, with attackers frequently targeting them in three of the reviewed articles ~\citep{king2020artificial,zouave2020artificially,guembe2022emerging}. Attackers often exploit URLs in various ways to achieve their malicious goals. By manipulating URLs, they can redirect users to malicious websites, steal sensitive information such as login credentials and personal data, or distribute malware.
\item Social media user profiles: Social media profiles were targeted in two sources ~\citep{king2020artificial,guembe2022emerging}. Attackers can use the profiles to hack accounts, spread misinformation, conduct phishing attacks or harvest personal information. 
\item Passwords: Passwords were a target in two sources~\citep{zouave2020artificially,guembe2022emerging}. Compromising passwords is a critical goal for attackers, as it provides access to user accounts and systems, potentially leading to broader breaches.
\item Details of systems: One article included targeting of details of systems~\citep{YAMIN2021102722}. This can include information about system configurations, network architectures, or software versions, which attackers can use to identify vulnerabilities and plan more effective attacks.
\end{itemize}

As can be seen, the same types of targets appear in many of the reviewed articles. These targets reflect the broad scope of cyberattacks and highlight the need for comprehensive security measures across different domains. 

\begin{table}
\caption{Attack targets found in the reviewed articles.}
\label{tab:targets}
\centering
\scriptsize
\begin{tabular}{p{1.35cm} p{1.25cm} p{1cm} p{1cm} p{0.8cm} p{0.8cm} p{0.8cm} p{0.8cm} p{0.8cm} p{0.8cm} c}
\toprule
Ref. & \cite{pistono2016unethical} & \cite{brundage2018malicious} & \cite{kaloudi2020ai} & \cite{king2020artificial} & \cite{truong2020artificial} & \cite{zouave2020artificially} & \cite{YAMIN2021102722} & \cite{wang2022poisoning} & \cite{guembe2022emerging} & $\sum$ \\
\midrule
\hangindent=0.5em
General data       &   & x &   &   & x & x & x &   &   & 4 \\
URLs                       &   &   &   & x &   & x &   &   & x & 3 \\
\hangindent=0.5em
Social media user profiles &   &   &   & x &   &   &   &   & x & 2 \\
Password                   &   &   &   &   &   & x &   &   & x & 2 \\
Systems                    &   &   &   &   &   &   & x &   &   & 1 \\
\bottomrule
\end{tabular}
\end{table}

\subsection{AI-driven attacks and targets}

We summarized various studies exploring the application of AI in malicious activities, highlighting different attack methods and targets.
Articles about AI hacking published from 2015 to 2018 did not report any attack methods, but targets were found, of which most could be categorized as data and sensitive data.
Articles in 2020 showed different AI methods and tools, such as GANs, CNN, RNN, LSTM, SVM, and SVC, aimed at attacking sensitive data, social media user profiles, passwords, and URLs.
Below, we present the articles in the same order as in Table~\ref{tab:targets}. 

Pistono and Yampolskiy focused on publishing papers related to malicious exploits and discusses the use of software with malicious capabilities, including truly artificially intelligent systems such as artificially intelligent viruses. The paper also introduces the term ``Hazardous Intelligent Software'' (HIS) to describe the use of intelligence in a malicious context. It highlights that intelligent systems can potentially become malevolent in various ways. However, the paper does not mention specific AI attack methods~\citep{pistono2016unethical}.

Brundage et al.\ provided a summary of workshop findings and the authors' conclusions on forecasting, preventing, and mitigating the detrimental impacts of malicious AI use. The targets included sensitive information or financial assets of individuals, specific members of crowds, and historical patterns of code vulnerabilities. The attacks were executed through various methods, such as spear phishing attacks, imitation of human-like behavior, facial recognition, the generation of custom malicious websites/emails/links, visual impersonation of another person in video chats, and the use of drones or autonomous vehicles to deliver explosives and cause accidents. Furthermore, the attackers were engaged in discovering new vulnerabilities and developing code to exploit them. However, no specific methods for these activities were mentioned in the report~\citep{brundage2018malicious}.

Kaloudi et al.\ investigate AI's threat to SCPS. It explores how AI can be used as a malicious tool, emphasizing its potential to increase attack speed and success rates. AI methods discussed include k-means clustering, RNN, LSTM, RL, and DNN. Case studies involve k-means clustering for phishing messages, RNN for deceptive reviews, LSTM for phishing URLs, RL for autonomous learning attacks, and DNN for cyberattacks. The paper also examines cyberattack methodologies, including DeepLocker, repurpose attacks, DeepHack, Deep-Phish, review attacks, and SNAP\_R~\citep{kaloudi2020ai}.

King et al.\ introduced the term ``AI-Crime'' (AIC) to address two key questions regarding the threats posed by AI in criminal activities and potential solutions to mitigate these threats. However, the article does not specify the methods employed in these AIC activities. The primary target of these activities is social media users, particularly through the use of phishing links~\citep{king2020artificial}.

Truong et al.\ provide an insightful overview of how artificial intelligence can be leveraged in cybersecurity, both for offensive and defensive purposes. They employ a diverse set of AI methods used in attacks, including SVM, RBM, MLP, KNN, CNN, PSO, GA, DAE, ANN, LS-SVM, NDAE, GWO, RWN, LFA, MLAPT, LSTM, and GLRT. The targets of these attacks encompass user identities, financial credentials, and sensitive data from large corporations, security agencies, and government organizations. These attacks serve various purposes, including detecting or categorizing malware, identifying network intrusions, countering phishing and spam attacks, mitigating Advanced Persistent Threats (APTs), and identifying domains generated by domain generation algorithms (DGAs)~\citep{truong2020artificial}.

Zouave et al.\ explored the possibilities and applications of AI throughout various stages of a cyberattack. The authors employ a wide range of artificial intelligence methods, including RNN, LSTM, NLP, GAN, KNN, logistic regression, SVC, decision tree, RF, gradient boosting regression tree, SVM, MLP, RLSC, OvA, CNN, and DGA. These attacks target URLs, individuals' personal data in search of relationships, passwords, captchas, and domains. The attacks are executed by creating deceptive URLs to evade automated detection, generating conversations that include harmful links and attachments, attempting password guessing and brute forcing, stealing passwords, solving captchas, and generating numerous random fake domains. The authors utilize cyberattack methodologies such as the DeepPhish algorithm, PassGAN, Torch RNN, Deeptcha, AGDs, and DeepDGA~\citep{zouave2020artificially}.

Yamin et al.\ turn their attention to raising awareness about the use of artificial intelligence as an attack method and assessed its impact on military operations. They employed GANs and Nash equilibrium to describe the attack methods. The targets of these attacks included traffic signs, medical image data, facial image data, digital recommendation systems, CT-scan data, speech and audio data, as well as network intrusion detection systems. The attacks were carried out using malicious AI algorithms designed to manipulate data to evade benign AI algorithm classifiers. The attack methodologies employed in these cyberattacks included DeepHack, DeepLocker, GyoiThon, EagleEye, Malware-GAN, UriDeep, Deep Exploit, and DeepGenerator~\citep{YAMIN2021102722}.

Wang et al.\ focus on poisoning attacks in machine learning, particularly within the context of automated vehicles. The authors utilize various attack methodologies harnessing AI techniques. These include deep learning and deep neural networks (DNN), known for their outstanding performance in recognition tasks like image classification and computer vision. Additionally, other methods are discussed, such as Generative Adversarial Networks (GAN), LSTM, SDN, DDQN (Deep Double Q-Network), Advantage Actor-Critic (A3C), SVM (Support Vector Machine), and singular value decomposition (SVD)~\citep{wang2022poisoning}.

Guembe et al.\ address the concern of AI-powered cyberattacks and further describe how AI can be used in such attacks for malicious goals. 
In the article, the authors also analyzed AI's performance in cyberattacks, finding that 56\% of AI-drive cyberattacks targeted the access and penetration phase, with 12\% in exploitation and command and control phases each, and 11\% and 9\% in reconnaissance and delivery phases. 
The result indicate that CNN was the most frequently used AI technique for access and penetration, followed by GAN and RNN, each used twice. Other techniques for malicious attacks that were demonstrated once include LSTM, SVC, SVM, cycle-GAN, TOD+CNN, RF, MLP, GBRT, KNN and DNN.
The targets of these attacks encompass public social media profiles, passwords, and URLs. The attacks are executed through techniques such as password guessing/cracking (brute-force attacks), intelligent captcha manipulation, smart abnormal behavioral generation, AI model manipulation, and the generation of fake reviews. The cyberattack methodologies employed by the authors include DeepLocker, DeepHack, PassGAN, and HashCat~\citep{guembe2022emerging}.

Finally, we turn to the two books identified by the screening process. Ward et al.\ describe the use of artificial intelligence as a new way to use technology for hackers and discuss ``poison'' attacks using machine learning algorithms. Furthermore, they consider automated vehicles and the potential for high-risk attacks on vehicle systems. However, during their discussion of AI hacking methods, no specific attack methods were mentioned~\citep{9694540}.
The use of AI has been identified as a cyberattack method and recognized as a potential risk. However, Clinton only presents AI as a hacking method, and we did not find any other specific attack methods~\citep{Clinton2022-zt}.

\section{Discussion}
\label{sec:discussion}

There are new attack strategies that cybercriminals continuously adapt, AI-driven techniques having recently been a focus in this regard. Our results indicate that primary targets (RQ2) include personal data as well as sensitive information held by governments, organizations, and individuals, spanning URLs, passwords, and critical systems. This emphasizes the significant threat posed by AI in the hands of malicious actors, as the ability to access and exploit such sensitive information can result in, e.g., identity theft, financial loss, and breaches of national security. 

The results also indicate that to achieve their goals (RQ1), cybercriminals have various AI methods at their disposal, which can be divided into categories: Classification, Regression, and Clustering. Each category encompasses several technologies utilized for different attack vectors. These methods do not differ from any other domain where AI is used, which indicates that any advances in technology can be adapted to red team use. 

\subsection{Offensive tactics}

There were several offensive tactics identified. Their usefulness and maturity levels vary, but the available technologies indicate how the threat landscape can evolve. 

\emph{Classification Techniques:} These methods include machine learning algorithms such as decision trees, CNN, RNN, LSTM, SVM, SVC, DNN, LS-SVM, NLP, OVA, DDQN, A3C, RLSC, and DGA.For instance, CNN and RNN can analyze patterns in data streams, while SVM and SVC excel at distinguishing between benign and malicious activities. Deep learning techniques, such as DNN and LSTM, can process vast amounts of data to uncover hidden insights that traditional methods might miss. Classification methods may help attackers to find new vulnerabilities and potential targets, since categorization of data can indicate patterns otherwise unknown. 

\emph{Regression Methods:} This category includes techniques such as GANs, RF, MLP, GBRT, MLP, ANN, Logistic Regression, and GLRT. Regression is used for prediction and estimation of variables, which in practice lead to threats such as password guessing and system security breaches. GANs, for example, are particularly effective in creating realistic but fake data, which can be used in phishing campaigns or to evade detection systems. RF and GBRT model complex relationships within data, providing cybercriminals with the predictive power needed to anticipate and exploit system weaknesses.

\emph{Clustering Strategies:} Attackers can also benefit from clustering methods such as k-means clustering, RBM, PSO, GA, DAE, and LFA. Clustering can be used for pattern analysis, creating possibilities for malicious activities. By grouping similar data points, attackers can uncover anomalies that signal potential vulnerabilities or valuable targets.

Attackers can use techniques such as the DeepPhish algorithm, PassGAN, Torch RNN, and Deeptcha. The mentioned tools and similar techinques facilitate attacks, including cracking passwords, phishing attacks, and infiltrating secure systems. The changing threat landscape should motivate the security research community, government agencies, and cybersecurity experts to be prepared for AI-based attacks. Red teaming using the AI-driven attacks could help the identification of vulnerabilities to novel techniques. The challenges created by AI cyberattacks necessitate the development of countermeasures to be effectively addressed.

\subsection{Defensive tactics}

While our findings highlight the evolving threats posed by AI-driven cyberattacks, defensive tactics and mitigation strategies to counter these threats are presented below. One promising approach is to use AI against AI, using it to identify and defend against cyber threats. Techniques such as anomaly detection, predictive analytics, and automated threat response systems can enhance the ability to detect and respond to malicious activities.

AI-based anomaly detection systems are used to inspect network traffic and user behavior for unsusual patterns indicating cyberattacks. By continuously learning from normal behavior, these systems can detect deviations that suggest potential security breaches, enabling intervention.
Machine learning models can be used to predict potential attack vectors and vulnerabilities. Predictive analytics can forecast where and how an attack might occur, allowing for preemptive measures to be put in place.

Implementing AI-driven automated threat response systems can significantly reduce the time between detecting and mitigating a threat. These systems can autonomously perform tasks such as isolating affected systems, applying patches, and reversing malicious actions, thereby limiting the damage caused by an attack.

Collaboration between organizations, government agencies, and cybersecurity researchers is essential. Sharing threat intelligence and best practices can help create a more robust defense against AI-driven cyber threats. Collaborative platforms and information-sharing frameworks can enhance collective resilience and response capabilities.
As AI-driven threats evolve, so must defensive strategies. Continuous monitoring, regular updates to security protocols, and adapting to new threat landscapes are essential to maintaining robust cybersecurity defenses.

\section{Conclusion}
\label{sec:conclusion}

The intersection of AI and cybersecurity presents both challenges and opportunities. 
The tactics and tools available to red team actors become more automated and capable of making decisions. 
In our study, we identified attack targets such as data, phony URL generation, social media profiles, password cracking and threats against systems. 
These targets, among others, can be attacked using methods that benefit from many of the well-known AI tools. 
These tactics and tools could be benefit red teams in their efforts to find new attack methods against targets similar to the ones found in this review. 
All stakeholders should stay informed about the latest advancements in AI-driven attack methodologies. 
Future research problems could include finding more attack targets used by AI attackers, and describing the various attack methods in more detail using existing taxonomies.

\section*{Acknowledgment}

The authors would like to thank Ms.\ Tuula Kotikoski for proofreading the manuscript.

\section*{Funding}

This research was partially funded by the Resilience of Modern Value Chains in a Sustainable Energy System project, co-funded by the European Union and the Regional Council of Central Finland (grant number J10052).

\bibliography{ai-cyberattacks-scoping-review-journal}

\end{document}